\documentclass[12pt,preprint]{aastex}

\newcounter{sub}
\newcounter{subeqn}[sub]
\newcommand\unit{{\mbox{\boldmath $e$}}}

\newcommand\be{\begin{equation}}
\newcommand\ee{\end{equation}}
\newcommand\lp{\left(}
\newcommand\rp{\right)}
\newcommand\ls{\left[}
\newcommand\rs{\right]}

\newcommand\st{\stepcounter{sub}}
\newcommand\stq{\stepcounter{subeqn}}
\newcommand\bea{\begin{eqnarray}}
\newcommand\eea{\end{eqnarray}}

\newcommand\xxi{{\mbox{\boldmath $\xi$}}}

\newcommand\nab{\mbox{\boldmath $\nabla$}}

\newcommand\B{{\bf B}}

\newcommand\J{{\bf J}}

\newcommand\no{\nonumber}

\received{2002 January 30}
\begin{document}

\title{R-Modes in the ocean of a magnetic neutron star
}
\author{
         Vahid Rezania 
}
\affil{Theoretical Physics Institute,
    Department of Physics,
    University of Alberta\\
    Edmonton, AB, Canada, T6G 2J1}
\affil{Institute for Advanced Studies in Basic Sciences,
          Zanjan 45195, Iran}

\begin{abstract}
We
study the dynamics of r-modes in the ocean of a magnetic neutron star.
We modeled the star's ocean with
a spherical rotating thin shell and assumed that the magnetic field
symmetry axis is not aligned to the shell's spin axis.
In the magnetohydrodynamic approximation, we calculate the frequency
of $\ell=m$ r-modes in the shell of an incompressible fluid.
Different r-modes with $\ell$ and $\ell\pm2$ are coupled by the
{\it inclined} magnetic field.
Kinematical secular effects for the motion of a fluid element in
the shell undergoing $\ell=m=2$ r-mode are studied.
The magnetic corrected drift velocity of a given fluid element
undergoing the $\ell=m$ r-mode oscillations is obtained.
The magnetic field increases the magnitude of the fluid drift produced
by the r-mode oscillations.
The drift velocity is strongly modulated by the inclined magnetic field.
We show that the magnetic field is distorted by the high-$\ell$ magnetic 
r-modes more 
strongly than by the low-$\ell$ modes.  Further, due to the shear produced 
by the r-mode drift 
velocity, the high-$\ell$ modes in the ocean fluid will damp faster 
than the low-$\ell$ ones. 

\end{abstract}


\section{Introduction}

In a series of papers \citet{BC95, BUC96} showed that neutron stars
accreting at high accretion rates $\gtrsim 10 ^{-9}$ M$_\odot$ yr$^{-1}$
are covered with massive oceans with densities $\gtrsim 10^9$ g cm$^{-3}$.
Both theoretical consideration of the nature of burning at these rates
\citep{FL87,Bild93,Bild95} and observational features of type I X-ray
bursts \citep{van95} reveal that the star burns the accreting hydrogen
and helium to mid-weight elements like C, O, Ne, etc.,  and accumulates
a massive ($\sim 10^{-6}$ M$_\odot$) degenerate liquid ocean.  Fusion
of this fuel to iron-group elements does not occur until densities
$\gtrsim 10^9$ g cm$^{-3}$.
The neutron stars with lower accreting rates (the atoll sources) burn
the accreted hydrogen and helium directly to iron-group elements via
the type I X-ray bursts which are crystallized by the larger Coulomb
force at much lower densities.
Further, they conjectured that the waves in these oceans might modulate
the outgoing X-ray flux at frequencies comparable to what is observed.

In this paper, motivated by Bildsten et al. conjecture we study the
evolution of r-modes in the ocean of magnetic neutron stars.
The r-modes those are analogous to the Rossby waves in the earth's
oceans, are driven by the Coriolis force in the rotating stars.
Their motions are dominantly toroidal and their oscillation frequencies
are proportional to the rotation rate of the star, $\Omega$.
Since even the small magnitude of Coriolis force drives r-modes in a 
rotating fluid, they should be considered in the dynamics of the fluid.
We model the star's ocean with a thin rotating shell of incompressible
inviscid fluid at the radius $R$ which is sandwiched between two hard
spheres. The role of spheres is to make sure that the fluid motion is
restricted to a two-dimensional spherical surface.

The paper is organized as follow.  In the next section, we set up
the magnetohydrodynamics equations for a spherical thin shell in
the background of a uniform magnetic field in which magnetic field
symmetry axis is not aligned with shell's spin axis.
We find the magnetic coupling coefficients and then the corrected
eigenfrequencies and eigenvectors for $\ell=m$ r-mode.
In section \ref{drift} we consider the kinematical secular drift of
r-modes, introduced by \citet{Rez00,Rez01a,Rez01b}, in a magnetic
ocean fluid.
We show that as well as the magnetic field strength causes the drift
magnitude to increase, the inclination of the magnetic field axis
modulates the r-mode trajectories.  The latter did not consider in
the previous studies.
Section \ref{discuss} is devoted for further discussions.

\section{R-modes in magnetic rotating shell}\label{mhd}

Historically, the study of oscillations of fluids in a rotating shell
in the presence of a magnetic field has been started about fifty years
ago by geophysicists.  Pioneering works by \cite{Hide66} and
\cite{Stew67}, are concerned with Earth's core magnetic modes that
may be accounted to the observed secular changes in the main geomagnetic
field at the Earth's surface.
It is well known that the Earth's magnetic field at the surface drifts
westward slowly, with the period of the order of one thousand years.
They assumed an unrealistic magnetic field configuration characterized
by a toroidal field of constant magnitude on the shell, and found the
magnetic corrected toroidal motions might cause the field drift at
the surface of the Earth.

In this paper we investigate the MHD perturbations of a uniformly
rotating thin spherical shell of incompressible fluid of radius $R$,
endowed with a uniform magnetic field (more realistic configuration)
which is given by
\st
\be\label{mag-1}
\B=B_p~(\cos\theta~ \unit_{\hat r}- \sin\theta~
        \unit_{\hat \theta})\,,
\ee 
on the shell.  
Since the shell is sandwiched between two spherical hard covers, the
motion of the fluid is restricted to a two-dimensional spherical surface.
Further, we assume the symmetry axis of the field makes an angle $\beta$
to the shell's rotation axis.
We work in an ideal MHD framework, so that the field lines are frozen
into the fluid, and the magnetic field rotates at the same rate as
the shell.

For an uniformly rotating shell of incompressible fluid with angular
frequency $\Omega$ around the $z$-axis, the fluid velocity perturbation field
in the corotating frame is exactly determined by
\st
\bea\label{xi-1}
\stq
&&\delta{\bf v}=\alpha\dot\xxi_{\ell m}\,,\\
\stq
&&\xxi_{\ell m}(\theta,\phi,t)=\frac{U_{\ell m}(R)}{R}\lp\frac{}{} -
   \partial_\phi Y_{\ell m}(\theta,\phi)/\sin\theta~\unit_{\hat \theta}
 +  \partial_\theta Y_{\ell m}(\theta,\phi)~\unit_{\hat \phi}
       \frac{}{}\rp~e^{-i\omega_{\ell m} t}\,,
\eea
for any arbitrary function $U_{\ell m}(R)$ and with dispersion relation
\stq \be 
\omega_{\ell
m}=-\frac{2m\Omega}{\ell(\ell+1)}\,.
\ee
Here $\alpha$ is the amplitude of the mode.
Equations (\ref{xi-1}) is the exact solution of the fluid equations of
motion in fully nonlinear regime, see \cite{LU01} for more detail
and references therein.

Recently, \cite{MR01} (MR) have studied dynamics of a rotating star
in the presence of a magnetic field in which its symmetric axis is
not aligned to the star's spin axis.
For the $\ell=m$ r-modes of an incompressible rotating fluid in the
background of uniform magnetic field they obtained a shift proportional
to the ratio of the magnetic energy to the rotational energy of the
star in the r-mode eigenfrequencies, only.

Following MR, we consider the magnetic force, produced by the fluid's
perturbations of the rotating shell, drives the motion of fluid elements.
As they've shown, the eigenvalue equation for eigenfrequencies and
eigenvectors of the normal modes of fluid in the presence of magnetic
field will be corrected by
\st\be\label{coeff-mag-1}
\omega_A \left( c_{A}- \frac{\cal M}{2\epsilon_A}
\sum_D \kappa_{AD} c_D \right) =\sigma_A c_{A},
\ee
where the $c_A$ and $\sigma_A$ are the magnetic corrected eigenvector
and eigenfrequency in the rotating frame, respectively.  Here $\epsilon_A$ is the energy 
of the mode in the rotating frame.  
The magnetic coupling coefficients, $\kappa_{AD}$, can be calculated
on the surface of sphere as
\st\bea\label{KAD}
&&\kappa_{AD}=
- \frac{1}{4\pi{\cal M}}
     \int \delta\B_A^*\cdot\delta\B_D
~ d\Omega\,,\\
\stq &&\delta\B_A=
\nab\times(\xxi_A\times\B)\,,
\eea
where ${\cal M}=B_p^2/4\pi$ is the magnetic field energy density
\footnote{
Equation (\ref{mag-1}) leads $\nabla
\times \B = 0$. As a result, the term containing $\J$ vanishes.
The term containing $\xxi\times\B$ does not vanish, but its contribution
is smaller by $\Omega^2R^2/c^2$ for r-modes and we neglect it in
equation \ref{KAD}.}.
Note that in the above calculations we expand any perturbation vectors
in terms of the modes of rotating star, $\xxi_A$, as
\st\be\label{xi-0}
\xxi=\sum_A c_A
\xxi_A e^{-i\sigma t}.
\ee
Here upper-case Latin subscripts are used to label solutions and
corresponds to the unique set of quantum numbers which describe the
solutions.

To calculate the magnetic coupling we use the method that is described
by MR.  They used the spin-weighted formalism in which:
(a)  the vectors are expressed in the orthonormal but complex basis
$\{\unit_+,\unit_-\}$ that is related to the usual orthonormal basis
$\{\unit_\theta,\unit_\phi\}$ by
\st\be
\unit_\pm = \frac{1}{\sqrt{2}} \lp \unit_\theta \pm \unit_\phi \rp\,;
\ee
(b) all functions are expanded by spin-weighted spherical harmonics,
$_s\!Y_{\ell m}(\theta,\phi)$, that are defined by \citep{Cam71}
\st\be
_s\!Y_{\ell m}(\theta,\phi) = \sqrt{\frac{(2\ell+1)}{4\pi}}
    d^\ell_{-sm}(\theta) e^{im\phi}
\ee
where the $d^\ell_{sm}(\theta)$ are the matrix representations for
rotations through an angle $\theta$ discussed in detail by \citet{Edm74}.
When $s=0$, the spin-weighted spherical harmonics reduce to the regular
spherical harmonics.
The orthogonality relations are
\st\be
\int_{4\pi} \;_s\!Y^*_{\ell'm'} \;_s\!Y_{\ell m} d \Omega
        = \delta_{\ell\ell'}\delta_{mm'}.
\label{ortho}
\ee
We use the convention that
$\;_\sigma\!Y^*_{\lambda \mu}(\theta,\phi) = (-1)^{\sigma+\mu}
\;_{-\sigma}\!Y_{\lambda -\mu}(\theta,\phi).$

A uniform magnetic field tilted by an angle $\beta$ from the shell's
spin axis can be written in spin-weight formalism as
\st\be
\B(\theta,\phi) = \sum_{\bar{m}=-1}^{\bar{m}=1}
       \lp\frac{}{}  
         b_0^{1,\bar{m}}(a)\;_0\!Y_{1\bar{m}}(\theta,\phi)\unit_0
     +   b_+^{1,\bar{m}}(a)\;_+\!Y_{1\bar{m}}(\theta,\phi)\unit_-
     +   b_-^{1,\bar{m}}(a)\;_-\!Y_{1\bar{m}}(\theta,\phi)\unit_+
       \frac{}{}\rp
\ee
where the functions $b_s^{1,\bar{m}}(a)$ are given by
\st\be
b_\pm^{1,\bar{m}} = \mp b_0^{1,\bar{m}}
      = \sqrt{\frac{4\pi}{3}} d^{1}_{\bar{m}0}(\beta),
\label{unib}
\ee
and the $d^{1}_{\bar{m}0}$ are given by
\st
\be
d^{1}_{10}(\beta) = \frac{1}{\sqrt{2}} \sin \beta, \quad
d^{1}_{00}(\beta) = \cos \beta, \quad
d^{1}_{-10}(\beta)= -\frac{1}{\sqrt{2}} \sin \beta.
\label{d1}
\ee
To rewrite equations (\ref{xi-1}) in spin-weighted formalism, at first
we normalize $\xxi$ such that the energy density of the modes in the
corotating system is $\epsilon_A={\cal T}$, where ${\cal T}$ is
rotational kinetic energy density of the shell. Therefore the r-modes
of non-magnetic incompressible thin shell have the form
\st
\begin{mathletters}
\bea
&& \omega_{\ell_A m_A} = - \frac{2m_A\Omega}{\ell_A(\ell_A+1)}\\
&& U_{\ell_A m_A} =
    R^2 \sqrt{ \frac{ \ell_A(\ell_A+1)}{4 m_A^2}}
\\
&& \epsilon_{\ell_A m_A} =  {\cal T} = \frac{1}{2} \rho R^2 \Omega^2\frac{}{}\,,
\eea
\end{mathletters}
The spin-weighted decomposition of the $A$th r-mode is
\st\be
\xi_A(x) =
  f^A_{+} \;_+\!Y_{\ell_Am_A}  \unit_- +
  f^A_{-} \;_-\!Y_{\ell_Am_A}  \unit_+ ,
\ee
where the functions $f_s^A(r)$ are given by (\cite{Sch01})
\st\be
f^A_+ = f^A_- =
 \frac{R}{2\sqrt{2}}\frac{\ell(\ell_A+1)}{m_A} \,.
\label{f:rmode}
\ee
Different components of the perturbed magnetic field,
$\delta\B_A$, due to the r-modes in the thin shell, can be found in
spin-weighted formalism as

\begin{footnotesize}
\st\bea\label{deltab}
\delta B_{A,s}(\theta,\phi)
     & = &\sum_{\lambda,\mu}  \delta B_{A,s}^{\lambda,\mu}
 \;_s\!Y^*_{\lambda \mu}\,,\\
\stq
\delta B^{\lambda,\mu}_{A,0}& =& \sum_{\bar{m}=-1}^{+1}
    C(\ell_A,1,\lambda)
    \left( \begin{array}{ccc}
        \lambda & \ell_A & 1\\
        \mu & m_A & \bar{m}
        \end{array}\right)
    \left( \begin{array}{ccc}
        \lambda & \ell_A & 1\\
        0   & -1 & 1
        \end{array}\right) \frac{1}{a} b^{1,\bar{m}}_0 f^A_+
    \left( 1 + (-1)^{\lambda+\ell_A} \right),\\
\label{deltab0}
\stq
\delta B^{\lambda,\mu}_{A,+}& =&
   (-1)^{\lambda+\ell_A+1} \delta B^{\lambda,\mu}_{A,-}\label{sym}  \\
&=&-\sum_{\bar{m}=-1}^{+1}
    C(\ell_A,1,\lambda)
    \left( \begin{array}{ccc}
        \lambda & \ell_A & 1\\
        \mu & m_A & \bar{m}
        \end{array}\right)
               \frac{1}{a} b^{1,\bar{m}}_0 f^A_+  \times \nonumber\\
\stq
&&\hspace{-.7cm}
      \left[
    \sqrt{ \frac{\ell_A(\ell_A+1)}{2}}
\left( \begin{array}{ccc}
        \lambda & \ell_A & 1\\
        -1  & 0 & 1
        \end{array}\right)
        +  \sqrt{ \frac{(\ell_A-1)(\ell_A+2)}{2}}
\left( \begin{array}{ccc}
        \lambda & \ell_A & 1\\
        -1  & 2 & -1
        \end{array}\right)
\right]\,,\no\\ \label{deltab+}
\eea
\end{footnotesize}

where $\left( \begin{array}{ccc}
\ell&k&\lambda\\
-s&-t&-\sigma \end{array}\right) $ is a Wigner 3-j symbol
\citep{Edm74} and the constant $C(\ell_A,\bar{\ell},\lambda)$ is
defined by
\st\be
C(\ell_A,\bar{\ell},\lambda) = (-1)^{\ell_A+\lambda+\bar{\ell}}
\sqrt{ \frac
{(2\ell_A+1)(2\bar{\ell}+1)(2\lambda+1)}{4\pi}}.
\ee
In equation (\ref{deltab}) $\lambda$ takes on values $\ell_A,
\ell_A\pm1$, although
Eq. (\ref{deltab0}) will survive only for $\lambda=\ell_A$.
Since the spin-weighted spherical harmonics obey the orthogonality
relation
(\ref{ortho}), integrals over all angles of quantities quadratic in
$\delta \B$ have the simple form
\st\bea\label{kappa1}
\kappa_{AD} &=& \int  \delta \B^*_A \cdot \delta \B_D d \Omega
    = \sum_{\lambda,\mu} \left(
      \delta B_{A,0}^{*\lambda,\mu} \delta B_{D,0}^{\lambda,\mu}
    + \delta B_{A,+}^{*\lambda,\mu} \delta B_{D,+}^{\lambda,\mu}
    + \delta B_{A,-}^{*\lambda,\mu} \delta B_{D,-}^{\lambda,\mu} \right),\\
\stq
&=&
\sum_{\lambda,\mu}\ls
\delta B_{A,0}^{*\lambda,\mu} \delta B_{D,0}^{\lambda,\mu}~\delta_{AD}
+\delta B^{*\lambda,\mu}_{A,+} \delta
B^{\lambda,\mu}_{D,+}
\left( 1 + (-1)^{\ell_A+\ell_D}\right)\rs,
\eea
where $\delta_{AD}$ is the Kronecker delta function.
The last step is due to the symmetry property (\ref{sym}).
It then follows that the magnetic coupling coefficients between
r-modes is zero unless both modes have the same parity. Since the
triangle inequalities
$\ell_A-1\le \lambda \le \ell_A+1$ and $\ell_D-1\le \lambda \le
\ell_D+1$ must be satisfied, only modes satisfying
$\ell_D = \ell_A, \ell_A\pm2$ have non-zero magnetic coupling.
Evaluating the terms appearing  in equations
(\ref{deltab0}) and (\ref{deltab+}) for $\ell_A=m_A$, the individual
terms for the allowed values of $\lambda$ are

\begin{footnotesize}
\st
\begin{mathletters}
\bea
\delta B^{\ell_A,\mu}_{A,0}&=&
(-1)^{\ell_A+1}
\sqrt{\frac{(2\ell_A+1)}{4}}~(\ell_A+1)
\sum_{\bar{m}=-1}^{+1} d^1_{\bar{m}0}
    \left( \begin{array}{ccc}
        \ell_A & \ell_A & 1\\
     -(\ell_A+\bar{m}) & \ell_A & \bar{m}
    \end{array}\right)\,,\\
\delta B^{\ell_A,\mu}_{A,+}&=&
(-1)^{\ell_A+1}
\sqrt{\frac{(2\ell_A+1)(\ell_A+1)}{8\ell_A}}
\sum_{\bar{m}=-1}^{+1} d^1_{\bar{m}0}
    \left( \begin{array}{ccc}
        \ell_A & \ell_A & 1\\
     -(\ell_A+\bar{m}) & \ell_A & \bar{m}
    \end{array}\right)\,,\\
\delta B^{\ell_A+1,\mu}_{A,+}&=&
(-1)^{\ell_A+1}
\sqrt{\frac{(\ell_A+1)(\ell_A+2)}{8\ell_A}}~\ell_A^2
\sum_{\bar{m}=-1}^{+1} d^1_{\bar{m}0}
    \left( \begin{array}{ccc}
        \ell_A+1 & \ell_A & 1\\
    -(\ell_A+\bar{m}) & \ell_A & \bar{m}
                \end{array}\right)\,,\\
\delta B^{\ell_A-1,\mu}_{A,+}&=&(-1)^{\ell_A+1}
\sqrt{\frac{(\ell_A-1)(\ell_A+1)}{8\ell_A}}~(\ell_A+1)^2
 \sum_{\bar{m}=-1}^{+1} d^1_{\bar{m}0}
    \left( \begin{array}{ccc}
        \ell_A-1 & \ell_A & 1\\
    -(\ell_A+\bar{m}) & \ell_A & \bar{m}
                \end{array}\right)\,.
\eea
\end{mathletters}
\end{footnotesize}

The magnetic coupling coefficients between different r-modes can now
be computed.  A straight-forward calculation yields the self-coupling
term, $\kappa_{AA}$ as
\st\be
\kappa_{AA} = -
      \frac{\ell_A+1}{4(2\ell_A+3)}
  \left\{
       \ell_A(\ell_A+1)(\ell_A+2)+3+
        \frac12 \ls \ell_A(\ell_A+1)(2\ell_A^2+2\ell_A-3)-9\rs
        \sin^2 \beta
  \right\}\,.
\label{final1}
\ee
The off-diagonal term, $\kappa_{A,A+2}$ and $\kappa_{A,A-2}$ will be
\st\bea
\label{final2}
\stq
&&\kappa_{A,A+2} = \kappa_{A+2,A}= -\frac18
         \sqrt{\frac{\ell_A(\ell_A+3)}{(2\ell_A+3)(2\ell_A+5)}}
          \ell_A(\ell_A+1)(\ell_A+3)^2\sin^2 \beta\,,\\
\stq
&&\kappa_{A,A-2} = \kappa_{A-2,A}= -\frac18
         \sqrt{\frac{(\ell_A-2)(\ell_A+1)}{(2\ell_A-1)(2\ell_A+1)}}
          (\ell_A-2)(\ell_A-1)(\ell_A+1)^2\sin^2 \beta\,,
\eea
Therefore, the equation of motion for mode's expansion coefficients
of the magnetically modified $A$th r-modes reduces to
\st\be
\sigma_{A}c_A = \omega_A
    \left[   \frac{\cal{M}}{2{\cal T}} |\kappa_{A,A-2}| c_{A-2}
           +  \lp 1 + \frac{\cal{M}}{2{\cal T}} |\kappa_{AA}|\rp c_A
           +  \frac{\cal{M}}{2{\cal T}} |\kappa_{A,A+2}| c_{A+2}
           \right]\,.
\label{final3}
\ee
The final equations (\ref{final1})-(\ref{final3}) give the magnetically corrected 
eigenfrequencies $\sigma_A$ and eigenvectors $c_A$ for the r-modes.
To compute these eigenfrequencies and eigenvectors one should solve
the $N$ dimensional linear algebraic system (\ref{final3}), where
$N$ is the number of the modes. Therefore, for a given $N=1,2,\cdots$,
the $N\times N$ matrix should be diagonalized.
Since it is not possible to solve an infinite dimensions system in
practice, we truncate the infinite matrix to a finite matrix and large
enough $N$.  
Further we note that for the lower inclination angles, the magnetic coupling 
coefficients $|\kappa_{AD}|$ increase by $\ell_A^3$, while for the larger 
inclination angles by $\ell_A^4$. Since we are working in the linear regime, 
we have to keep $({\cal M}/2{\cal T})|\kappa_{AA}|<1$.   
The ratio of magnetic energy density to rotational energy density in
the covered ocean at the surface of star is
\st\be\label{BT}
\frac{{\cal M}}{2{\cal T}}= 8\times 10^{-3}
    \lp\frac{B}{10^{13}{\rm~ G}}\rp^2
    \lp\frac{10^{9}{\rm~ g\;cm}^{-3}}{\rho}\rp
    \lp\frac{10{\rm ~km}}{R}\rp^2
    \lp\frac{10^3 {\rm ~Hz}}{\Omega}\rp^2\;,
\ee
where $\rho$ is the ocean's density.
In tables \ref{t:table1} and \ref{t:table2} we present the eigenfrequencies 
of
$\ell_A=m_A=1, 2, \cdots, 6$ r-modes.  In table \ref{t:table1} we assume the
inclination angle between field axis and shell's spin axis is zero,
$\beta=0$, and calculate the eigenfrequencies $\sigma_A$ for different
ratio of the magnetic energy density to the rotational energy density,
${\cal M}/2{\cal T}$, by increasing the magnetic field strength, see 
equation (\ref{BT}).
In table \ref{t:table2} we compute the eigenfrequencies by increasing the 
inclination
angle for a specific energy ratio ${\cal M}/2{\cal T}=10^{-2}$. As a 
result, the eigenfrequency of the mode increases by increasing both
magnetic field energy density and inclination angle.

In the next section we consider the kinematical secular effects of r-modes in the oceanic fluid.

\section{The r-mode drift by the magnetic field}\label{drift}

In a series of papers \citet{Rez00,Rez01a,Rez01b} showed the existence
of kinematical secular velocity field of r-mode oscillations which interacts
with the background magnetic field of the star.  Although, this
interaction is non-linear and should be considered in non-linear regime,
one can find some second-order quantities from linear result \citep{Rez00}.
As an example, the second-order secular toroidal drift which appear
on isobaric surfaces, can be obtained through the linear velocity field.

To investigate this secular effect in the fluid motion undergoing the
magnetic corrected r-mode oscillations in the ocean fluid, we
consider the real part of the magnetic corrected fluid velocity of
$\ell_A=m_A$ r-mode, equations (\ref{xi-1}),
\st
\bea\label{v1}
\stq
&&{\dot\theta} (t)=-\sum_{A}~\alpha ~Q(\ell_A)~c_{A}~{\hat \sigma}_A~
(\sin\theta)^{\ell_A-1}\cos(\ell_A\phi-2\pi{\hat \sigma}_A t),\\
\stq
&&{\dot\phi} (t)=\sum_{A}~\alpha ~Q(\ell_A)~c_{A}~{\hat \sigma_A}~
\cos\theta~(\sin\theta)^{\ell_A-2}\sin(\ell_A\phi-2\pi{\hat \sigma_A} t),
\eea
where ${\hat \sigma}_A=\sigma_A/\Omega$ and
\stq\be
Q(\ell_A)=\lp-\frac12\rp^{\ell_A} \frac{1}{\ell_A !}
\sqrt{\pi\ell_A(\ell_A+1)(2\ell_A+1)!/4}.
\ee

In equations (\ref{v1}) $\sigma_A$ and $c_A$ are the magnetic corrected eigenfrequency
 and eigenvector, respectively, and the time, $t$, is in unit of the star's period, $P$.

In figures 1 to 3 we plotted the result of numerical integration of
equations (\ref{v1}) in the shell's northern hemisphere for the magnetic
corrected $\ell_A=m_A=2$ r-mode with amplitude $\alpha=0.1$.  Note,
we assumed that at $t=0$ the only $\ell_A=m_A=2$ r-mode is present in
the fluid and all calculations are done in the corotating frame.

In the case of zero inclination angle, $\beta=0$, our results for all
mode's latitudes are in good agreement to those results which have
been reported by \citet{Rez00,Rez01a}.
In figure 1 we plotted $\theta(t)$ vs $\phi(t)$ from $t=0$ to $t=7.45
P$ (almost 5 cycles of the mode), with initial values
$\phi(0)=0,~\theta(0)=\pi/2$, $\ell_A=m_A=2$, zero inclination angle
$\beta=0$, and two different values of ${\cal M}/2{\cal T}=0$ and $10^{-1}$.
The aligned field causes the drift in $\phi$-direction (produced by
the r-mode) to increase.
The corrected expression for the drift velocity of a given fluid element
in the background of the uniform magnetic field will be
\footnote{
Following \citet{Rez00,Rez01a} equation (\ref{v-d}) is obtained by
expanding equations (\ref{v1}) in powers of $\alpha$, averaging over
a gyration, and retaining only the lowest-order nonvanishing term.}
\st 
\bea\label{v-d}
&&{\bf v}_{\rm d}= \sum_A K_A(\theta)\alpha(t)\sigma_A(t)\unit_{\phi}\,,\\
\stq
&&K_A(\theta)=Q^2(\ell_A)c^2_A(\sin\theta)^{2\ell_A-4}
\ls{\over} \sin^2\theta-2(\ell_A-1)\cos^2\theta{\over}\rs\,.
\eea
As a result, because of nonzero inclination angle, the drift velocity
of a fluid element undergoing the $\ell_A$ r-mode oscillation is affected by all 
$\ell_A\pm 2$ r-mode oscillations. 
Equation (\ref{v-d}) reduces to one obtained by 
\citet{Rez00,Rez01a}, for the non-magnetic $\ell_A=m_A=2$ r-mode, ie.  
${\bf v}_{\rm d}= K_2(\theta)\alpha(t)\omega_2(t)\unit_{\phi}$.
Therefore, the total displacement in $\phi$ from the onset of the
oscillation at $t_0$ to time $t$ is
\st
\be\label{phi-d}
\Delta\phi(\theta,t)
=\sum_A K_A(\theta)\int_{t_0}^t\alpha^2(t')\sigma_A(t')dt'\,.
\ee
The change in
$\phi$ due to the magnetic field is given in tables \ref{t:table3} and 
\ref{t:table4}. As a result,
$\Delta\phi=\phi_{\rm mag}-\phi_{\rm nonmag}$ increases by increasing
the magnetic energy density.  Further the high-$\ell_A$ magnetic r-modes are 
drifted more than the low-$\ell_A$ ones. 

As shown in figures 2 and 3, the r-mode trajectories in both $\theta$-
and $\phi$-directions are strongly modulated by the {\it inclined} magnetic
field.
In figure 2 from left to right, we plotted time evolution of r-mode
displacements in $\theta$-direction, in $\phi$-direction, and the
projected ($\theta,~\phi$) r-mode trajectory, respectively.
All panels correspond to
${\cal M}/2{\cal T}=10^{-2}$ and are plotted with initial conditions
$\phi(0)=0,~\theta(0)=\pi/2$.
The inclination angle is $\beta=0$ for the top, $\beta=\pi/6$ for
the middle, and $\beta=\pi/2$ for the bottom panels.
The larger the field's angle the stronger the modulation.
First and second columns of figure 2 show how
$\theta$- and $\phi$- displacements of the initial $\ell_A=m_A=2$
r-mode are affected by the inclined field, respectively.
In the third column we show the motion of fluid element in the
northern hemisphere of the shell undergoing the $\ell_A=m_A=2$ r-mode
oscillations.
The projected trajectories
$\theta(t)\sin\theta(t)\cos\phi(t)$ and
$\phi(t)\sin\theta(t)\cos\phi(t)$ show how the inclined magnetic
field change the motions which will be detected by a  
corotating observer in the rotation equator of the shell.

It is interesting to note that the results show the modulations 
are more sensitive to the field inclination angle rather
than the magnetic field strength.  This can be understood from equations
(\ref{v1}).
A non-diagonal matrix of the corrected eigenvectors $c_A$
causes modulations in both $\theta$ and $\phi$ trajectories.  
For any non-zero value of ${\cal M}/2{\cal T}\ne 0$, no matter
how small, and $\beta\ne 0$ the eigenvectors matrix is not diagonal.
Furthermore, the magnitude of the corrected eigenvector $c_A$ changes
significantly by varying $\beta$, while  
it is almost {\it constant} for different (and non-zero)
values of ${\cal M}/2{\cal T}$.
This fact is shown in figure 4.  In the left panel,
we plotted the element $c_2(2,2)$ as a function of $\beta$ for a 
fixed value of
${\cal M}/2{\cal T}\approx 10^{-10}$.  As is shown, 
by increasing $\beta$ the value of $c_2(2,2)$ drops by
$\sim .02$.   While, the change in the element $c_2(2,2)$ due to the 
${\cal M}/2{\cal T}$ for a fixed value $\beta=\pi/2$ is~ 
$\lesssim 10^{-8}$, which is shown in the right panel of figure 4.
Therefore,
no matter how the value of the energy ratio small, the fluid motions 
are strongly modulated by a magnetic field making a large inclination angle 
to the star's symmetry axis. 

Furthermore, the larger the fluid element's latitude, the stronger
the modulation.
This feature is shown in figure 3 which is plotted by different 
initial conditions $\phi(0)=0,~\theta(0)=\pi/6$ for both $\beta=0$
(the top panels) and $\beta=\pi/2$ (the bottom panels). Comparing
both bottom panels in figures 2 and 3 reveals that the magnetic field
(having the same energy density and inclination angle) has less effect
on the lower latitude modes.

Finally we note that $\Delta\phi$ also increases by increasing the inclination angle,
 see the third column of figures 2 and 3.  Therefore, the motion of a fluid element of
 the shell undergoing the magnetic r-mode oscillation is drifted more than nonmagnetic 
r-mode oscillation both by field energy density and field inclination angle.  These
 properties can be understood by considering both $\Delta\phi$ and ${\bf v}_d$ are
 proportional to the corrected eigenfrequency and eigenvector, see equations (\ref{v-d})
 and (\ref{phi-d}).  The corrected eigenvector changes by changing inclination angle while 
the corrected eigenfrequency increases by increasing field energy density.

The toroidal fluid motions produced by r-mode are perpendicular to the magnetic field.  
They distort the field and increase the field energy density. As shown by \citet{Rez00} 
if the energy that it needs to distort the magnetic field required by the 
mode is greater than the mode energy during an r-mode oscillation, the 
distortion of the magnetic field which is proportional to the ${\bf v}_d$, will prevent 
the oscillation from occurring.  To make an estimate at which field strength the distortion 
effect might get to be important we calculate the ratio of the change in the magnetic energy 
density of the shell, $\delta E_m=\int(\delta B^2/4\pi)d\Omega$, to the mode energy density 
during one oscillation as
\st\bea\label{dB-energy}
\frac{\delta E_m}{\epsilon_A}&\approx&\frac{{\cal M}}{2{\cal T}}|\kappa_{AA}|
\,,\\
\stq
&\approx& 10
    \lp\frac{B}{10^{15}{\rm~ G}}\rp^2
    \lp\frac{10^{9}{\rm~ g\;cm}^{-3}}{\rho}\rp
    \lp\frac{10{\rm ~km}}{R}\rp^2
    \lp\frac{10^3 {\rm ~Hz}}{\Omega}\rp^2
    \left\{\begin{array}{l}
      \ell_A^3~~~~~{\rm for~small~\beta}\\
      \ell_A^4~~~~~{\rm for~large~\beta}
      \end{array}\right.\;.
\eea
Thus $\delta E_m>\epsilon_A$ if $B>B_{\rm critical}\approx 3\times 10^{14}
\rho_9^{1/2}
R_{10}\Omega_3 \ell_A^\eta$, where $\eta=-3/2$ and $\eta=-2$ for small and large  
inclination angles, respectively.  Here $\rho_9=\rho/10^9$ g cm$^{-3}$, $R_{10}=R/10$ km, 
and $\Omega_3=\Omega/10^3$ s$^{-1}$. 
Hence the smaller field strength would prevent the high-$\ell_A$ r-mode 
oscillations.  Further, equation (\ref{dB-energy}) shows that for the higher field 
inclination angle the distortion effect will be stronger. 

In addition the increase in the energy of the magnetic field produced by 
the r-mode 
drift reduces the mode energy and causes damping.  
The rate that the magnetic field extracts energy density from the mode can be 
estimated by 
\st\be\label{B-phi}
(1/4\pi)\int B_\phi(dB_\phi/dt)d\Omega\approx \sum_A K_A(\theta)\alpha^2\sigma_A B_pB_\phi\,.
\ee 
The azimuthal field $B_\phi$ is generated by r-mode drift velocity ${\bf v}_d$ from the 
background poloidal field.  The change in $B_\phi$ can be calculated approximately for 
the magnetic r-modes from the time $t_0$ that the oscillation begins to time $t$ by $\Delta 
B_\phi(t)\approx\sum_A\int_{t_0}^t K_A(\theta)\alpha^2(t)\sigma_A(t)B_p dt$ \citep{Rez00}.   
Equation (\ref{B-phi}) shows that the energy loss from the mode increases
when either the field energy density or the inclination angle is increased.
Furthermore, since the high-$\ell_A$ 
magnetic r-modes are drifted more than the 
low-$\ell_A$ ones, they will damp faster than the low-$\ell_A$ modes.

\section{Discussion}\label{discuss}

In this paper, by assuming neutron stars in the low-mass X-ray binaries (LMXBs) are 
covered by degenerate massive oceans, we have studied the interaction of the inclined 
magnetic field with r-mode oscillations of an incompressible ocean fluid in a thin shell 
approximation.
Our analysis shows that in the case of large inclination angle,
even for {\it weak} magnetic field energy density
in comparison with the star's rotational energy density, $\ell$ and
$\ell\pm2$ r-modes are coupled significantly.

The kinematical secular drift produced by the r-mode, is studied in
the background of the inclined magnetic field.   The results show that
the magnetic field tends to increase the fluid drift in the
$\phi$-direction.  As a result, for the same time period, the fluids
in the presence of the magnetic field undergo larger drift than fluids
with no background magnetic field.  

Since the magnetic field is distorted by the r-mode, we showed that the field distortion 
by the high-$\ell$ magnetic r-mode is stronger than the low-$\ell$ modes.  
Thus a smaller critical field is needed to prevent the high-$\ell$ r-mode 
oscillation 
from occurring.  
Further, the shear produced by the drift velocity of the magnetic r-mode  
will damp high-$\ell$ modes faster than the low-$\ell$ ones. 
Note that in the present paper, we assume the r-mode is stable,
so we do not deal with its unstable growth in the magnetic star as has
been considered by \cite{Rez00}.  

Observational and phenomenological studies of more than twenty LMXBs reveal that these 
accreting neutron stars are source of millisecond oscillations, burst oscillations, and 
quasi-periodic oscillations in Hz-kHz range.  Different authors proposed different possible 
mechanisms to explain the observations in LMXBs, but the regularity of these oscillations 
in many sources led the investigators to consider radial or nonradial oscillations in a 
coronal accretion flow near the Eddington limit, see the review by \cite{Van00}.  
By assuming neutron stars in LMXBs covered by an ocean fluid \citet{BC95, BUC96} showed 
that both low and high radial orders of ocean g-modes might be responsible for observed 
low frequency ($\sim$ Hz) oscillations in these stars.
Recently \cite{Heyl01} discussed different possible oscillations (the g-modes, the Kelvin 
modes, and the buoyant r-modes) that may excite in the neutron star ocean during a Type-I 
burst, and concluded that the buoyant r-modes may meet the observations of low frequency 
oscillations in Type-I X-ray bursts quite well.

Figures 2 and 3 show how the inclined magnetic field modulates $\ell_A=m_A=2$ 
r-modes. The associated envelope frequency, $f_{\rm envelope}$, decreases as 
$f_{\rm envelope}\sim \nu_s/10 - \nu_s/16$ when the inclination angle increases 
from $\beta\sim\pi/6-\pi/2$.  Here $\nu_s=\Omega/2\pi$ is the star's spin frequency
.  The frequencies of these new modes are comparable to those observed oscillations 
in the LMXBs in low frequencies.  For the star's spin frequency $\nu_s\approx 300$ Hz, 
the wave envelope frequency is $f_{\rm envelope}\sim 20-30$ Hz.  Different frequencies 
may be obtained by different initial $\ell_A=m_A$ r-modes.
Further, the amplitude of
the modulated oscillations increases by increasing the inclination
angle and/or mode's
altitude.

Finally we note that during the evolution of the r-mode instability, 
one would expect the growth of the amplitude of the new modes.  
This feature can be understood from equations (\ref{v1}) which show that 
both $\theta$ and $\phi$ trajectories of the r-mode
are proportional to the mode's amplitude, $\alpha$.
Further, changes in the star's spin frequency in this period,
lead to change in the frequency of modulated modes as well.
In figure 5 we plotted time evolution of $\theta(t)$-displacement
of the $\ell_A=m_A=2$ r-mode, from $t=0$ to $t=100P$, for different
initial mode's amplitudes
$\alpha=0.1,~0.5$ (the top panels) and $\alpha=0.8,~0.9$ (the bottom panels).
All panels corresponds to ${\cal M}/2{\cal T}=10^{-2}$, $\beta=\pi/2$ and
are plotted with initial conditions $\phi(0)=0$ and $\theta(0)=\pi/2$.
By increasing $\alpha$ both amplitude and period of the modulated
oscillations increase.  Further, for $\alpha$ close to the
saturation limit, $\alpha_{\rm sat}=1.0$, the behavior of the modulated
oscillations changes drastically.
Therefore, to analyze the problem carefully,
one needs to solve both the nonlinear equations (\ref{v1}) and the
equations governing the r-mode evolution together.  This is currently
under investigation \citep{Rez02}.

Though there is still no firm observational evidence for the low
magnetic field strength
in the LMXBs, these objects are believed to be neutron stars with
weak magnetic fields
($B\lesssim 10^9$ G), high accretion rates
(${\dot M}\gtrsim 10^{-9}$ M$_\odot$ yr$^{-1}$),
and milliseconds periods \citep{Van00}.   
Our calculations show the existence of an inclined (even {\it weak})
magnetic field modulates
the normal modes of an ocean fluid to a new mode with frequency $f_{\rm 
envelope}$.
Therefore, by observing normal modes in the neutron star's
ocean, one would expect to observe these new modes.

\acknowledgments

The author is grateful to S. M. Morsink, S. Sengupta and L. Rezzeolla
for reading the manuscript and numerous discussions.  I would like to 
thank referee for his useful comments.
This research was supported by the Natural Sciences and Engineering
Research Council of Canada.

\clearpage

\begin{deluxetable}{lllllll}
\tablecaption{
The eigenfrequency of $\ell_A=m_A$ r-modes in the thin shell in the
presence of a uniform magnetic field, $\sigma_A/\Omega$.  The angle
between magnetic field axis and shell's spin axis is $\beta=0$.
The energy in the magnetic field is
${\cal M} = B_p^2/4\pi$ and the rotational energy is ${\cal T} = \rho
a^2\Omega^2/2$.
A number $a\times 10^{\pm b}$ is written as $a\pm b$.
\label{t:table1}}
\tablehead{
\colhead{${\cal M}/2{\cal T}$}&\colhead{$\ell_A=1$}
&\colhead{$\ell_A=2$}&\colhead{$\ell_A=3$}
& \colhead{$\ell_A=4$}&\colhead{$\ell_A=5$}
&\colhead{$\ell_A=6$}
}
\startdata
 0.0 &-.10000+01&-.66667+00&-.50000+00&-.40000+00&-.33333+00&-.28571+00\\
 1.0-04 &-.10000+01&-.66686+00&-.50004+00&-.40056+00&-.33415+00&-.28684+00\\
 1.0-03 &-.10005+01&-.66943+00&-.50838+00&-.41797+00&-.37886+00&\\
 1.0-02 &-.10090+01&-.68595+00&-.53500+00&-.45591+00& &\\
\enddata
\end{deluxetable}

\clearpage

\begin{deluxetable}{llll}
\tablecaption{
The eigenfrequency of $\ell_A=m_A$ r-modes in the thin shell in the
presence of a uniform magnetic field, $\sigma_A/\Omega$ for different
value of inclination angle, $\beta$.  The ratio of magnetic energy to
the rotational energy is
${\cal M}/2{\cal T}=10^{-2}$.
A number $a\times 10^{\pm b}$ is written as $a\pm b$.
\label{t:table2}}

\tablehead{
\colhead{$\beta$}&\colhead{$\ell_A=1$}&\colhead{$\ell_A=2$}
&\colhead{$\ell_A=3$}
}
\startdata
 0.0 &-.10090+01&-.68595+00&-.53500+00\\
$\pi/6$&-.10080+01&-.68874+00&-.54775+00\\
$\pi/4$&-.10068+01&-.69038+00&-.55827+00\\
$\pi/3$&-.10056+01&-.69155+00&-.56837+00\\
$\pi/2$&-.10043+01&-.69250+00&-.57834+00 
\enddata
\end{deluxetable}

\clearpage

\begin{deluxetable}{ll}
\tablewidth{10cm}
\tablecaption{
Change in displacement $\phi$,
$\Delta\phi=\phi_{mag}-\phi_{nonmag}$ for
$\ell_A=m_A=2$ r-mode, due to the magnetic field, from $t=0$ to
$t=7.45 P$.
Initial values $\phi(0)=0,~\theta(0)=\pi/2$ and zero inclination
angle $\beta=0$ are considered.
A number $a\times 10^{\pm b}$ is written as $a\pm b$.
\label{t:table3}}
\tablehead{
\colhead{${\cal M}/2{\cal T}$}&\colhead{$\Delta\phi$}
}
\startdata
 0.0 & 0.0 \\
 1.0-04 & .36403-06\\
 1.0-03 & .19299-05\\
 1.0-02 & .29707-03\\
 1.0-01 & .10390-01\\
\enddata
\end{deluxetable}

\clearpage

\begin{deluxetable}{ll}
\tablewidth{10cm}
\tablecaption{
Change in displacement $\phi$,
$\Delta\phi=\phi_{mag}-\phi_{nonmag}$ for different 
$\ell_A=m_A$ r-mode, due to the magnetic field, from $t=0$ to
$t=7.45 P$.
Initial values $\phi(0)=0,~\theta(0)=\pi/2$, ${\cal M}/2{\cal T}=10^{-2}$ 
and zero inclination angle $\beta=0$ are considered.
A number $a\times 10^{\pm b}$ is written as $a\pm b$.
\label{t:table4}}
\tablehead{
\colhead{$\ell_A=m_A$}&\colhead{$\Delta\phi$}
}
\startdata
 1 & .56340-06 \\
 2 & .29707-04\\
 3 & .39012-03\\
 4 & .14556-02\\
\enddata
\end{deluxetable}

\clearpage

\begin{figure}
\includegraphics{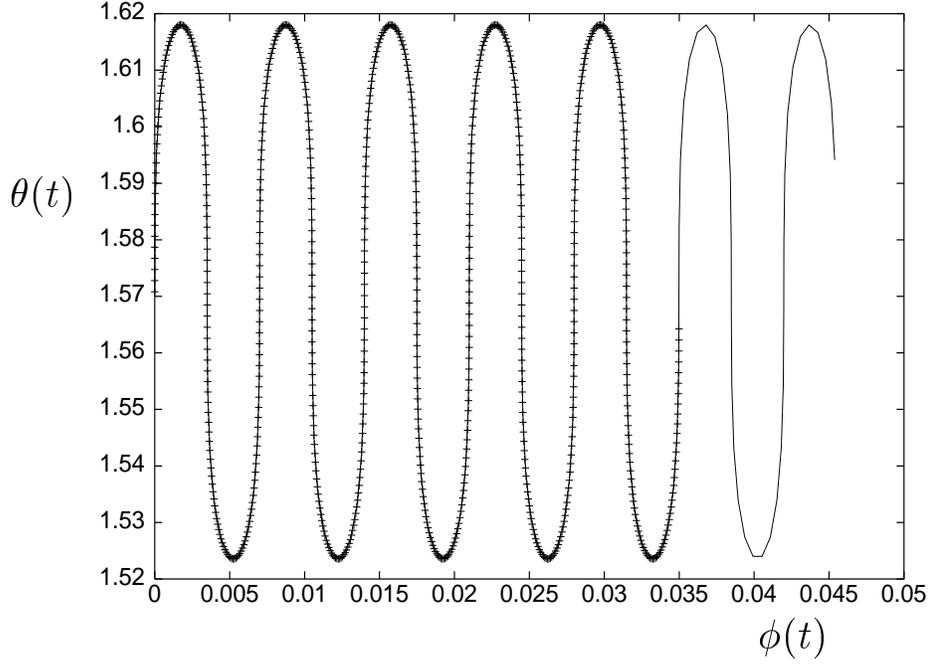}
\null
\vspace{9cm}
\caption{
Motion of a fiducial fluid element in the northern hemisphere of a
shell undergoing
$\ell_A=m_A=2$ r-mode with amplitude $\alpha=0.1$.
The projected trajectories are plotted for 7.45 of period of star
(almost 5 cycles of the mode), with initial values
$\phi(0)=0,~\theta(0)=\pi/2$ and zero inclination angle
$\beta=0$. The dash sign refers to zero magnetic field, ${\cal
M}=0$, while the continuous line corresponds to assumed value
${\cal M}/2{\cal T}=10^{-1}$.  Both $\theta$ and $\phi$ are in radian.
\label{fig1}}
\end{figure}

\clearpage

\begin{figure}[h]
\vbox to2.6in{\rule{0pt}{2.6in}}
\includegraphics{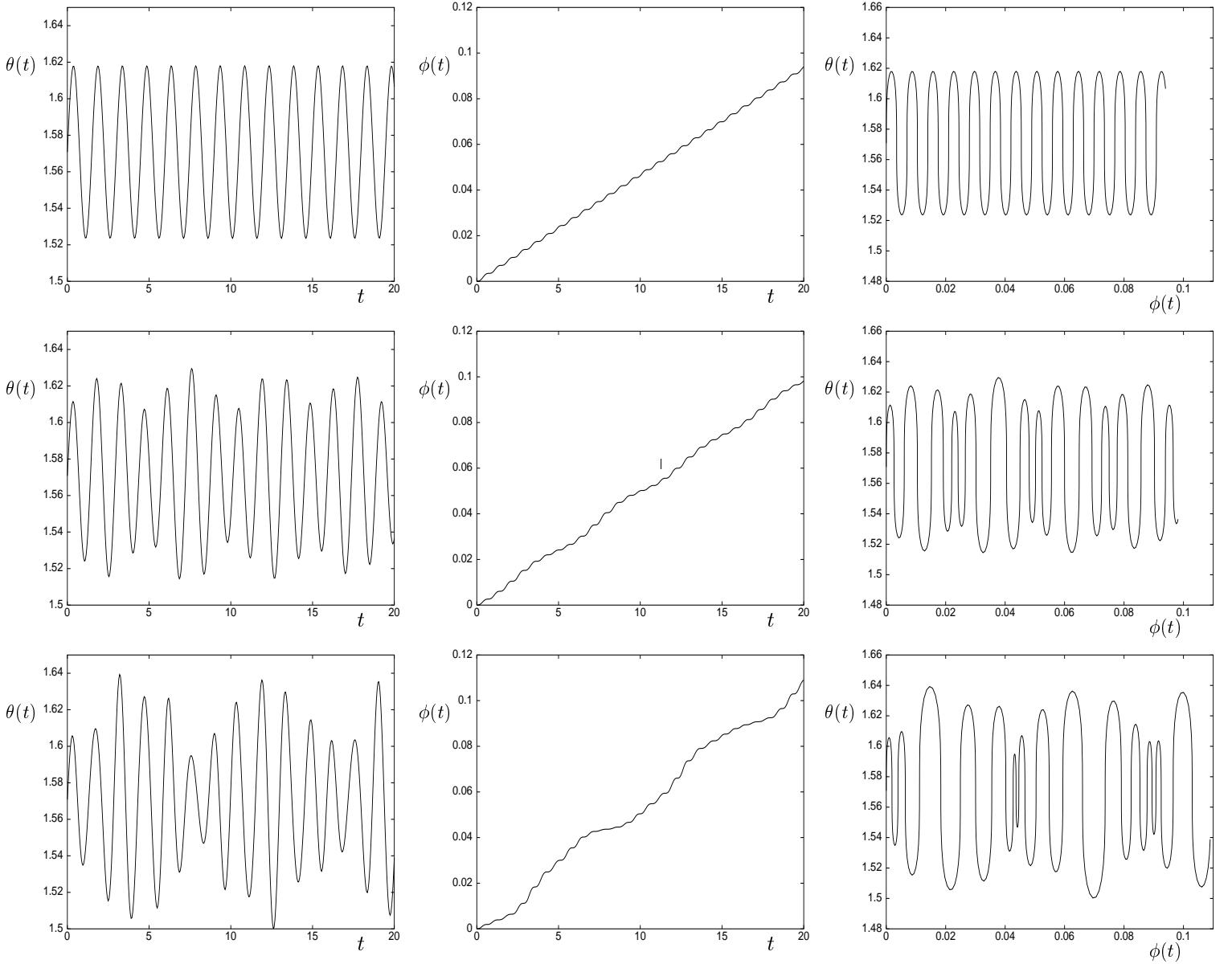}
\null
\vspace{11cm}
\caption{Motion of a fiducial fluid element in the northern hemisphere
of a shell undergoing $\ell_A=m_A=2$ r-mode with amplitude $\alpha=0.1$.
The projected trajectories are plotted for 20 of period of star
(almost 13 cycles of the mode), with initial values
$\phi(0)=0,~\theta(0)=\pi/2$ and ${\cal M}/2{\cal T}=10^{-2} $. Both
$\theta$ and $\phi$ are in radian and $t$ is in unit of star's period, $P$.
(a) The top panels are plotted for $\beta=0$, (b) The middle panels
are plotted for $\beta=\pi/6$, (c) The bottom panels are plotted for
$\beta=\pi/2$.}
\end{figure}

\clearpage

\begin{figure}[h]
\vbox to2.6in{\rule{0pt}{2.6in}}
\includegraphics{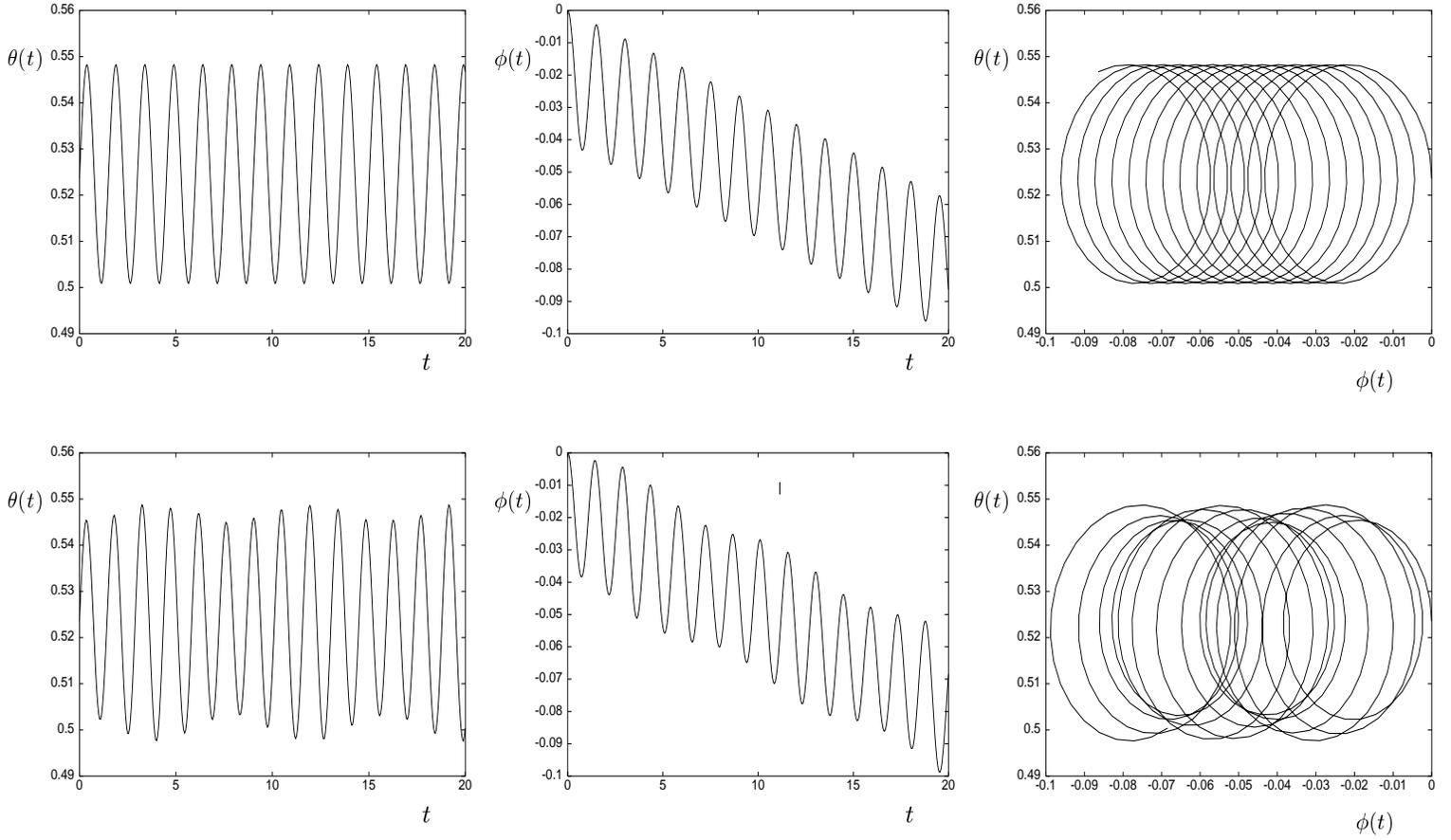}
\null
\vspace{11cm}
\caption{Same as figure 2 with initial values $\phi(0)=0,~\theta(0)=\pi/6$.
(a) The top panels are plotted for $\beta=0$,
(b) The bottom panels are plotted for $\beta=\pi/2$.}
\end{figure}

\clearpage

\begin{figure}[h]
\vbox to2.6in{\rule{0pt}{2.6in}}
\includegraphics{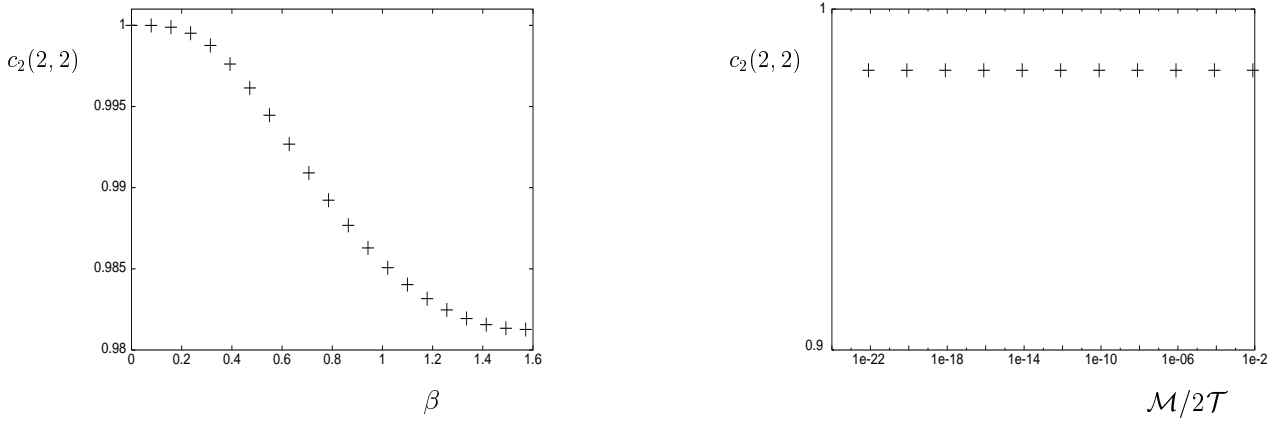}
\null
\vspace{11cm}
\caption{
Change in the element $c_2(2,2)$ as a function of the inclination angle  
$\beta$ and/or the energy ratio ${\cal M}/2{\cal T}$.
(a)The left panel is plotted for ${\cal M}/2{\cal T}\approx 10^{-10}$.  
The values of $\beta$ in radian. 
(b) The right panel is plotted for $\beta=\pi/2$. Both axes have 
logarithmic scale.
}
\end{figure}

\clearpage

\begin{figure}[h]
\vbox to2.6in{\rule{0pt}{2.6in}}
\includegraphics{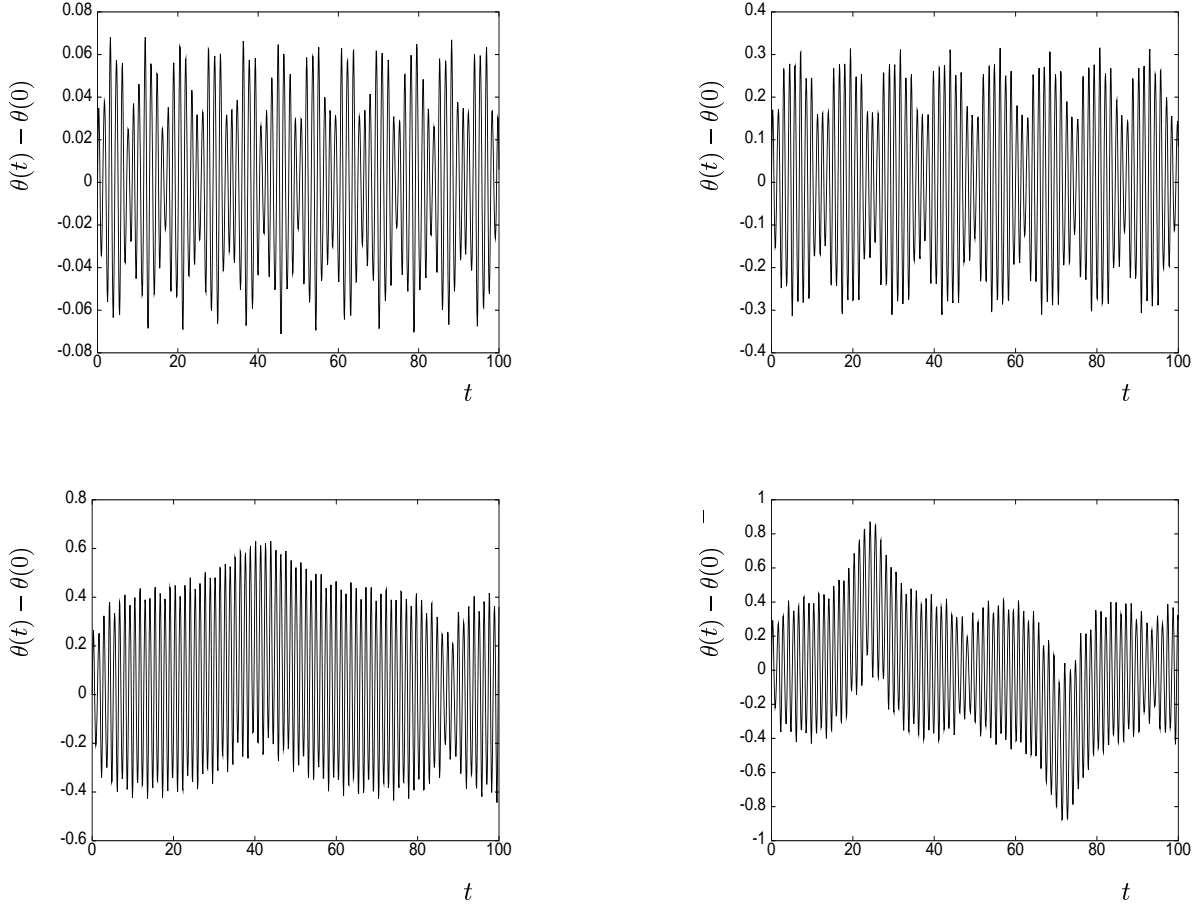}
\null
\vspace{11cm}
\caption{
Time evolution of $\theta(t)-\theta(0)$ displacement
of the $\ell_A=m_A=2$ r-mode, from $t=0$ to $t=100P$, for different
initial mode's amplitudes $\alpha$.
(a) The left-top panel is plotted for $\alpha=0.1$,~
(b) The right-top panel is plotted for $\alpha=0.5$,~
(c) The left-bottom panel is plotted for $\alpha=0.8$,~
(d) The right-bottom panel is plotted for $\alpha=0.9$. 
All panels correspond to ${\cal M}/2{\cal T}=10^{-2}$, $\beta=\pi/2$ and
are plotted with initial conditions $\phi(0)=0$ and $\theta(0)=\pi/2$.
}
\end{figure}

\end{document}